\documentstyle[aps,prb,epsf]{revtex}

\begin{document}

\draft

\twocolumn[\hsize\textwidth\columnwidth\hsize\csname@twocolumnfalse%
\endcsname

\title{Skyrmion lattice melting in the quantum Hall system}
\author{Carsten Timm\cite{CTpresent} and S. M. Girvin}
\address{Department of Physics, Indiana University, Bloomington,
Indiana 47405}
\author{H.A. Fertig}
\address{Department of Physics and Astronomy, University of
Kentucky, Lexington, Kentucky 40506}
\date{April 6, 1998}

\maketitle

\begin{abstract}
The melting and magnetic disordering of the skyrmion lattice in the
quantum Hall system at filling factor $\nu\approx 1$ are studied. A
Berezinskii-Kosterlitz-Thouless renormalization group theory is
employed to describe the coupled magnetic and translational degrees
of freedom. The non-trivial magnetic properties of the skyrmion
system stem from the in-plane components of the non-collinear
magnetization in the
vicinity of skyrmions, which are described by an antiferromagnetic
$XY$ model. In a Coulomb gas formulation the `particles' are the
topological defects of the $XY$ model (vortices) and of the lattice
(dislocations and disclinations). The latter frustrate the
antiferromagnetic order and acquire fractional vorticity in order to
minimize their energy. We find a number of melting/disordering
scenarios for various lattice types. While these results do not
depend on a particular model, we also consider a simple classical
model for the skyrmion system. It results in a rich
$T=0$ phase diagram. We propose that the triangular and square
skyrmion lattices are generically separated by a centered rectangular
phase in the quantum Hall system.
\end{abstract}

\pacs{PACS numbers: 73.40.Hm, 75.40.Cx, 75.70.Ak}

]

\section{Introduction}

\subsection{General Remarks and Motivation}
\label{sec.intro}

For nearly two decades the study of the quantum Hall effect has been
one of the most productive fields of condensed matter
physics.\cite{QHrev1,QHrev2,Chak,QHrev3,Das} Recently, quantum Hall
systems with additional degrees of freedom have received considerable
attention.\cite{QHex1,QHex2,QHex3,QHex4,Moon,Pelle,DSZ,QHferro} In the
simplest case this degree of freedom is the electron spin. Ideas
developed for this system can be adapted for other multi-component
quantum Hall systems such as coupled layers, wide quantum wells, and
quantum wells in semiconductors with several degenerate conduction
band minima.\cite{QHferro}

Here, we study the effect of the electron spins. We are motivated by
recent nuclear magnetic resonance\cite{Barrett} and specific
heat\cite{Bayot} measurements exhibiting interesting finite
temperature spin physics. At the Landau level
filling factor $\nu=1/m$, where $m$ is an odd integer, the ground
state of the two-dimensional electron gas is a strong
ferromagnet,\cite{Sondhi,QHferro,Cote97} {\it i.e.}, the electronic
spins are completely aligned even in the limit of vanishing Zeeman
coupling. Perhaps surprisingly, the effective Zeeman field in this
system is rather small because of band structure effects.
In the following we consider the case $\nu\sim1$.
The low-energy excitations of the system at $\nu=1$
are spin waves gapped at the Zeeman energy. However, the quantum Hall
ferromagnet also has topologically non-trivial
excitations, which are (not strictly correctly) referred to as
{\it skyrmions\/}\cite{Lee,Sondhi,Moon} in analogy with the Skyrme
model in nuclear physics.\cite{Skyrme,BP,Raja} They can be thermally
created in pairs of vanishing topological charge, similar to vortices
in two-dimensional superfluids. Skyrmions are
in fact also present in conventional itinerant ferromagnets such as
iron but do not seem to have any observable consequences at low
temperatures. What makes them crucial for
quantum Hall ferromagnets is that the vanishing
diagonal conductivity $\sigma_{xx}=0$ together with the finite Hall
conductivity $\sigma_{xy}=\nu e^2/h$ makes the skyrmions carry
a quantized electrical charge of $\pm\nu e$.\cite{Sondhi,Moon}
As a result, skyrmions (or antiskyrmions) are present even in the
ground state if we move slightly away from
$\nu=1$.\cite{QHground,Slattice} (For our purposes skyrmions and
antiskyrmions behave identically and we refer to both as
``skyrmions.'')

If one dopes electrons or holes into the
two-dimensional electron gas at $\nu=1$, they enter the system as
skyrmions with charge $\mp e$ but with more than one flipped spin.
This effect can be seen in measurements of the
magnetization as a function of filling factor.\cite{Barrett}
The result is a non-collinear ground state since the magnetization in
the vicinity of the skyrmion centers has components perpendicular to
the magnetic field, which have a vortex-like configuration. In a
collinear magnet the SO(3) spin symmetry is broken to a SO(2) symmetry
with respect to rotations
around the magnetic field direction. It has one Goldstone mode, which
is gapped at the Zeeman energy in the presence of a magnetic field.
In a non-collinear magnet the SO(2) symmetry is further broken
and there are two Goldstone modes, only one of which is gapped.
The other, gapless mode corresponds to rotation of the non-collinear
spin configuration around the magnetic field axis.
Thus non-collinearity leads to the appearance of a new low-energy
SO(2) $\sim$ U(1) degree of freedom.
In the long-wavelength limit the orientation
of the in-plane components of the magnetization of
a skyrmion can be described by a single U(1) phase factor $e^{i\phi}$
or by the angle $\phi$.

Moving further away from $\nu=1$, more and more skyrmions are present
and their interaction becomes important. The skyrmion interaction
contains a repulsive, long-range Coulomb part and a short-range
contribution related to the U(1) degree of freedom. The latter
term, which we here call the {\it magnetic\/} interaction, favors
antiparallel alignment of the U(1) ``spins.''
Brey {\it et al}.\cite{Slattice} recognized that the magnetic
interaction could lead to a square lattice of skyrmions instead
of the usual triangular (hexa\-go\-nal) lattice since the square
lattice allows N\'eel order of the U(1) degree of freedom whereas
magnetic order is frustrated on the triangular lattice. Since the
magnetic interaction is of short range, the
Coulomb interaction dominates for small skyrmion densities and one
expects a triangular crystal. The U(1) degree of freedom is then
frustrated, with neighboring angles $\phi$ differing by
$\pm120^\circ$. On the other hand, if the density is
sufficiently high, the energy gained from N\'eel ordering on the
square lattice may outweigh the lost Coulomb energy.
Further lattice types may also be possible, {\it e.g.},
a centered rectangular lattice, {\it i.e.}, a square lattice stretched
along the $(11)$ direction, with N\'eel order.
In Sec.~\ref{sec.class} we employ a simple classical model of the
skyrmion system to illustrate which lattice types and
structural transitions may be expected. We find a surprisingly rich
phase diagram for the classical ground state.
%% NEW
The classical ground state has also been investigated by  %SMG
Rao {\it et al}.\cite{Rao} employing a non-linear sigma model
and by Green {\it et al}.\cite{Green}, who also study the
lattice dynamics, see Sec.~\ref{sec.class} below.   %SMG
%% END NEW

If the skyrmion positions were fixed to the ideal lattice at all
temperatures, the long-wavelength physics, in particular the critical
properties, would be well described by an antiferromagnetic lattice
$XY$ model. We would then expect a Berezinskii-Kosterlitz-Thouless
(BKT) tran\-si\-tion,\cite{BKT,Kost74} which separates a
low-temperature phase of bound pairs of logarithmically interacting
vortices and antivortices from a high-temperature phase where
large pairs are broken in the sense that their interaction is
completely screened. These broken pairs, which essentially consist
of free vortices and antivortices, destroy quasi-long-range order.
In the skyrmion lattice, however, the positions are not fixed and
the lattice itself can melt.

The critical properties of a two-dimensional lattice without any
internal degree of freedom have been
successfully described by Nelson and Halperin\cite{NH}
and by Young\cite{Young} applying BKT
theory to dislocations and disclinations of the lattice.
Melting of the triangular and square lattices proceeds
in two steps, both well described by BKT theory,\cite{NH}
unless one or both of these transitions is preempted by a first-order
melting transition. At the lower
transition bound pairs of dislocations with opposite Burger's vector
decouple, leading to a liquid crystal phase with short-range
translational order but persisting quasi-long-range orientational
order of nearest-neighbor bonds. Note that dislocations are, like
vortices, topological defects with logarithmic bare interaction.
For the triangular lattice, the liquid crystal
phase is called {\it hexatic\/} because is shows quasi-long-range
order with respect to a sixfold rotational symmetry, whereas for
the square lattice it is called {\it tetratic\/}
(fourfold symmetry).\cite{NH} However, the square lattice is unstable
in a system with Coulomb repulsion as the only interaction. At the
higher transition temperature, pairs of disclinations, {\it i.e.},
defects in the bond orientation field, unbind, leading to an isotropic
fluid. The {\it bare\/} disclination interaction is confining but the
presence of free dislocations above the lower melting temperature
leads to a logarithmic interaction.\cite{NH}

For the skyrmion lattice, the U(1) degree of freedom may not only
stabilize the square lattice structure,\cite{Slattice} we also expect
the short-range magnetic interaction to be strongly affected
by lattice deformations, leading to the coupling of magnetic and
lattice degrees of freedom. There are two aspects to this coupling:
First, the low-energy collective modes ($XY$ spin waves and lattice
vibrations) may be coupled. From
general arguments\cite{magphon,Cote97,HS}
the dispersion of the lattice vibration mode, usually called
{\it magnetophonon\/} mode, close to the Brillouin zone center
is expected to have the form $\omega\propto k^{3/2}$.
C\^ot\'e {\it et al}.\cite{Cote97} have performed time-dependent
Hartree-Fock calculations for the collective mode spectrum of square
skyrmion crystals. The authors indeed find two distinct low-energy
branches. One is linear in ${\bf k}$ and is interpreted as the
gapless $XY$ spin wave mode, whereas the other has the $k^{3/2}$
magnetophonon dispersion. There is no sign of mixing of these
two modes at small $k$. In Sec.~\ref{sec.dyn} we briefly show
that the classical skyrmion model reproduces these features.

\begin{figure}
\centerline{\epsfxsize 7cm\epsfbox{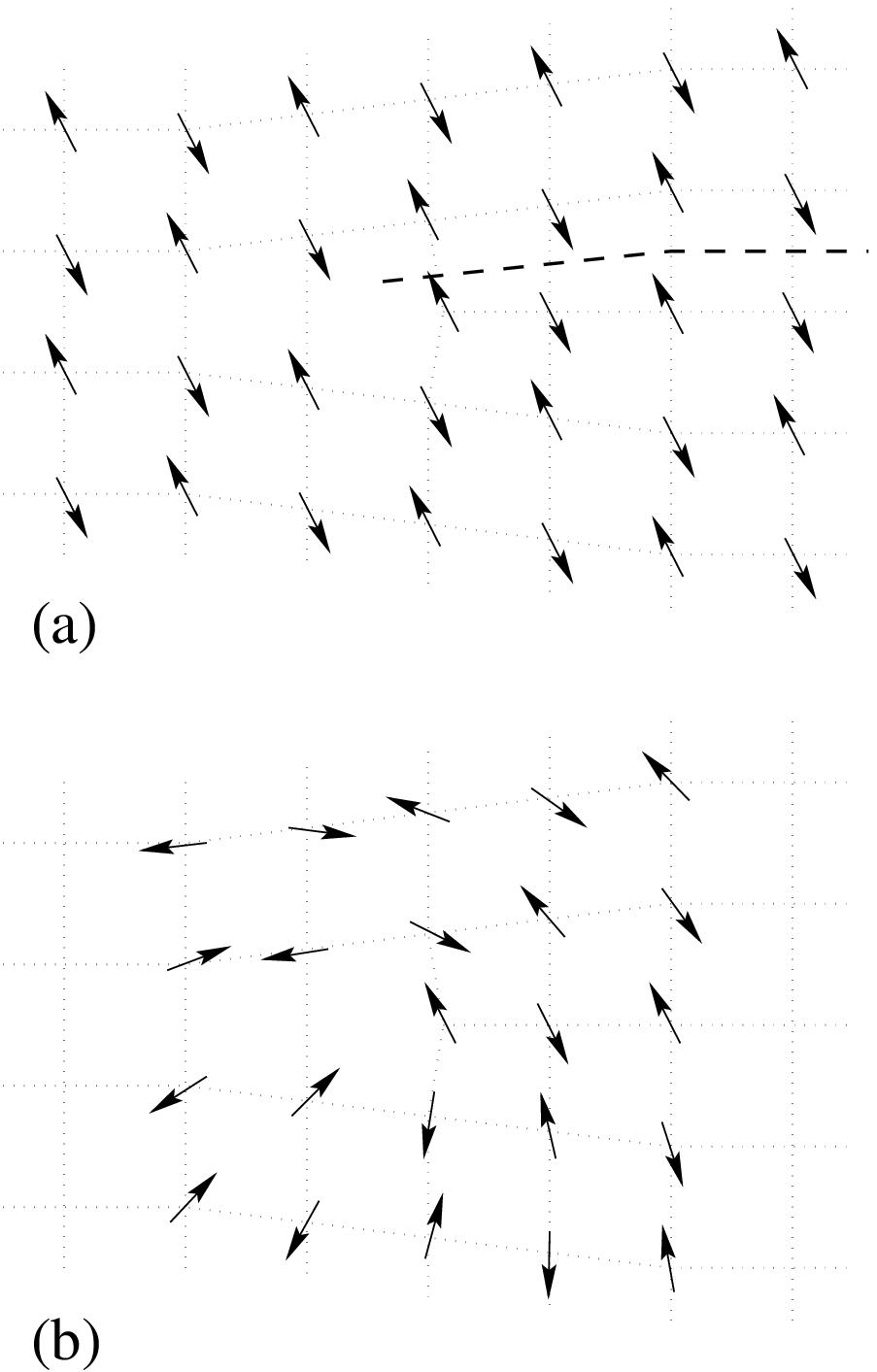}}
\caption{Sketch of a dislocation in a square skyrmion lattice.
The arrows denote the internal U(1) degree of freedom. As seen
in (a), dislocations lead to a phase mismatch of $\pm\pi$ in
the U(1) degree of freedom. In (b) the U(1) angles have been allowed
to relax and the dislocation has acquired half a vortex.}
\label{fig.Smis}
\end{figure}

Second, despite the fact that the collective modes are largely
decoupled, the topological excitations (vortices,
dislocations, and disclinations)
may be coupled, leading to an interplay of the magnetic BKT
transition and the BKT melting transitions. That something
non-trivial happens is easily seen from Fig.~\ref{fig.Smis}: A
dislocation in a square lattice leads to a phase mismatch of $\pm\pi$
in the U(1) degree
of freedom since the nearest-neighbor coupling is antiferromagnetic.
Naively one could expect that this mismatch along the dashed line in
Fig.~\ref{fig.Smis} leads to a linear, confining term in the
interaction of a dislocation pair. However, the magnetization can
relax so as to minimize the mismatch energy, as shown in
Fig.~\ref{fig.Smis}(b). In this relaxed configuration, the dislocation
has acquired {\it half\/} a U(1) vortex
and the dislocation interaction is again logarithmic. The main
objective of this paper is to illustrate this point further and to
explore its consequences for melting and magnetic disordering of
various skyrmion lattice types.

The vorticity acquired by dislocations can be derived using a gauge
theory of elasticity following ideas of Cardy {\it et al}.\cite{Cardy},
who have studied a soft square lattice of
antiferromagnetically coupled Ising spins.
Dislocations frustrate the antiferromagnetic order is this system
as well. In Sec.~\ref{sec.gauge} we sketch the gauge theory for
the case of a square skyrmion lattice.

Experimentally, the situation is less clear. The presence of
skyrmions for $\nu\approx1$ but $\nu\neq1$ is firmly established by
magnetization measurements by Barrett {\it et al}.\cite{Barrett},
using optically pumped nuclear magnetic resonance (NMR) techniques.
Less certain,
the very large nuclear relaxation rate $T_1^{-1}$ seen in this
regime\cite{Barrett} is interpreted in terms of the gapless $XY$
magnon mode.\cite{Cote97}
This mode couples strongly to the nuclear spins
because of its large $S^{x,y}$ components and its gaplessness.
This coupling opens a channel for rapid spin-lattice relaxation
of nuclear spins. In these experiments\cite{Barrett} the skyrmions are
probably usually in a liquid state. Nevertheless the gapless $XY$
mode is presumably still present as an overdamped mode.

Bayot {\it et al}.\cite{Bayot} find a strongly enhanced specific
heat $C$ for $\nu\approx1$, which suggests strong
coupling between electronic and nuclear spins so that the large
specific heat of the nuclear spins is in fact measured. Again, a
plausible coupling mechanism is provided by the gapless $XY$
magnons. The filling factor dependence of $C$ is consistent with this
picture.\cite{Bayot} The temperature
dependence of $C$ shows a sharp peak at very low $T$. This peak
may indicate a skyrmion lattice melting transition. Our
so far quite speculative interpretation is the following:
Neglecting the skyrmions for a moment,
the Zeeman energy of nuclear spins within the quantum well
containing the two-dimensional electron gas is (Knight) shifted
because of their coupling to the polarized electron gas. Outside
of the quantum well there is no such Knight shift and the mismatch
in the Zeeman energy prevents the spins within
and outside of the well from coming into thermal equilibrium.
If skyrmions are present in a liquid state
they move around, leading to motional narrowing and an averaged,
but still finite, Knight shift within the well.
In a lattice state, however, there are regions around the
skyrmions where the electronic magnetization is perpendicular to
the external field and the Knight shift vanishes. The nuclear
spins outside of the well can come into equilibrium with the nuclei
in these regions. Hence, specific heat
measurements suddenly see the nuclei outside of the quantum well
when a skyrmion lattice forms. Below the transition the apparent
specific heat drops off again, which may indicate that the coupling
is strong only in the vicinity of the transition where critical
slowing down causes the electronic motion time scale to pass
through the NMR time scale. To our knowledge, these
experiments\cite{Bayot} are the only ones showing signs of
a finite-temperature phase transition in a single-layer
quantum Hall system. Recent experiments
using resonant inelastic light scattering off
{\it double-layer\/} systems\cite{Pelle} also show signs of a
finite-temperature transition.\cite{Pelle,DSZ}

The objective of this paper is to illustrate several of the points
raised above, in particular we wish to explore the consequences of
the vorticity acquired by lattice defects for the melting and
magnetic disordering transitions of skyrmion lattices. In
Sec.~\ref{sec.KT} we study these transitions for several possible
lattice types. We introduce
a generalized BKT renormalization group theory for a Coulomb gas
with more than one species of particle carrying more than one kind
of charge. The specific lattice types we
study are motivated by the possible ground states of the
simple classical skyrmion model of Sec.~\ref{sec.class}. It should be
kept in mind, however, that the BKT theory does not depend on any
particular model.

\subsection{Gauge Theory of Elasticity\\
for the Skyrmion Lattice}
\label{sec.gauge}

Following Cardy {\it et al}.\cite{Cardy} we here formulate a
gauge theory for the magnetic and elastic energy of a square
skyrmion lattice, {\it i.e.}, a soft square lattice with an
$XY$ degree of freedom. Other skyrmion lattice types can be treated
similarly. Let $S({\bf R}) \equiv e^{i\phi({\bf R})}$ be
the $XY$ spin field, where ${\bf R}$ is a lattice vector.
We define a magnetic order parameter
\begin{equation}
\Phi({\bf R}) \equiv S({\bf R})\, e^{i\pi(r_x+r_y)/a} ,
\end{equation}
where ${\bf r}({\bf R})$ is the actual position of the skyrmion
belonging to the ideal lattice vector ${\bf R}$ and $a$ is the lattice
constant. This picture breaks down in the presence of free
disclinations, {\it i.e.}, above the upper melting transition,
because then the $x$ and $y$ components of the position
vector ${\bf r}$ are no longer well-defined.
The additional phase factor turns the spins on one sublattice
through $\pi$, thereby mapping the antiferromagnet onto a ferromagnet.
In the presence of dislocations this is not possible:
The phase factor is no longer unique and
$\Phi({\bf R})$ cannot be both continuous and single-valued.
Using a continuum notation, the exchange energy is
\begin{equation}
H_{\text{ex}} = \int d^2\!R\, \rho\,
  (\mbox{\boldmath$\nabla$}\Phi)^\ast (\mbox{\boldmath$\nabla$}\Phi) .
\end{equation}
Next, we define two translational order parameters
\begin{eqnarray}
\Psi_x({\bf R}) & \equiv & e^{i2\pi r_x/a} ,  \\
\Psi_y({\bf R}) & \equiv & e^{i2\pi r_y/a} .
\end{eqnarray}
Although ${\bf r}({\bf R})$ is not continuous and single-valued in
the presence of dislocations, the fields $\Psi_{x,y}$ are.
Continuum elastic theory yields the elastic energy of a
square lattice,\cite{elast}
\begin{equation}
H_{\text{el}} = \int d^2\!R \left(
  \mu\, u_{ij}u_{ij} + \frac{\lambda}{2}\, u_{ii}u_{jj}
  + \nu\, u_{xx}u_{yy} \right)
\end{equation}
where summation over repeated indices is implied and
\begin{equation}
u_{ij} \equiv \frac12 \left( \frac{\partial r_i}{\partial R_j}
  + \frac{\partial r_j}{\partial R_i} \right) .
\end{equation}
The last term in the elastic energy would be absent for a
triangular crystal or an isotropic medium.\cite{elast}
Expressing the $u_{ij}$ in terms of derivatives of the fields
$\Psi_{x,y}$ and integrating by parts, the elastic energy becomes
\begin{eqnarray}
H_{\text{el}} & = & \int d^2\!R \,\frac{a^2}{8\pi^2} \bigg(
  \mu\, (\mbox{\boldmath$\nabla$} \Psi_i)^\ast
      (\mbox{\boldmath$\nabla$} \Psi_i)  \nonumber \\
& & {}+ (\mu+\lambda)\,\Big[ (\partial_x\Psi_x)^\ast(\partial_x\Psi_x)
  + (\partial_y\Psi_y)^\ast(\partial_y\Psi_y) \Big]  \nonumber \\
& & {}+ (\mu+\lambda+\nu)\,\Big[ \Psi_x(\partial_x\Psi_x)^\ast
      \Psi_y^\ast(\partial_y\Psi_y)  \nonumber \\
& & \quad{}+ \Psi_y(\partial_y\Psi_y)^\ast
      \Psi_x^\ast(\partial_x\Psi_x) \Big] \bigg) .
\label{Helast}
\end{eqnarray}
The long-range Coulomb repulsion in the skyrmion lattice
drives the Lam\'e coefficient $\lambda$ to infinity---the lattice
is incompressible.\cite{Th78,FHM} Thus only the first term on the
right-hand side of Eq.~(\ref{Helast}) is relevant
(the other terms may introduce constraints on $\Psi_{x,y}$, which
we ignore in the following since they do not affect our
argument).

A translation in the $x$ ($y$)
direction leads to phase factors in $\Psi_x$ ($\Psi_y$) and in
$\Phi$. A spin rotation leads to a phase factor in $\Phi$ alone.
We can express the symmetries under translation and spin rotation
as {\it gauge symmetries\/}: We introduce three two-component gauge
fields ${\bf A}_0$, ${\bf A}_x$, and ${\bf A}_y$ and write the energy
$H\equiv H_{\text{ex}}+H_{\text{el}}$ as
\begin{eqnarray}
H & = & \int d^2\!R\, \bigg[ \rho
   \left|\left(\mbox{\boldmath$\nabla$}-i{\bf A}_0
    -\frac{i}{2}{\bf A}_x-\frac{i}{2}{\bf A}_y\right) \Phi\right|^2 
\label{1H1} \\
& & {}+ \frac{a^2\mu}{8\pi^2}\,
    |(\mbox{\boldmath$\nabla$}-i{\bf A}_x)\Psi_x|^2
      + \frac{a^2\mu}{8\pi^2}\,
    |(\mbox{\boldmath$\nabla$}-i{\bf A}_y)\Psi_y|^2
  \bigg] .  \nonumber
%%% label in first line
\end{eqnarray}
This Hamiltonian is invariant under gauge transformations with respect
to any of the three gauge fields:
\begin{eqnarray}
{\bf A}_0 & \to & {\bf A}_0 + \mbox{\boldmath$\nabla$}\theta_0 ,
  \quad \Phi \to e^{i\theta_0}\,\Phi , \\
{\bf A}_x & \to & {\bf A}_x + \mbox{\boldmath$\nabla$}\theta_x ,
  \quad \Phi \to e^{i\theta_x/2}\,\Phi ,
  \quad \Psi_x \to e^{i\theta_x}\,\Psi_x , \! \\
{\bf A}_y & \to & {\bf A}_y + \mbox{\boldmath$\nabla$}\theta_y ,
  \quad \Phi \to e^{i\theta_y/2}\,\Phi ,
  \quad \Psi_y \to e^{i\theta_y}\,\Psi_y . \!
\end{eqnarray}
These three transformations correspond to spin rotation, translation
in the $x$ direction, and translation in the $y$ direction,
respectively.
The {\it matter fields\/} $\Phi$, $\Psi_x$, and $\Psi_y$ in the
Hamiltonian, Eq.~(\ref{1H1}), are only coupled through the gauge
fields.

We now discuss the topological defects in this theory.
Magnetic vortices, {\it i.e.}, vortices in $\Phi$, are threaded by
one flux quantum with regard to ${\bf A}_0$.
Dislocations correspond to topological defects in $\Psi_x$ or
$\Psi_y$, depending on the Burger's vector orientation.
The elementary defect in, say, $\Psi_x$ is a unit vortex. It is
threaded by one flux quantum in ${\bf A}_x$. This seems to make the
field $\Phi$ multi-valued since its phase changes by $\pi$ if one
moves around the vortex. However, if the $XY$ spin part $S$ in $\Phi$
itself contains {\it half\/} a vortex (or antivortex) the order parameter
$\Phi$ is again single-valued and continuous. This corresponds to
$\pm1/2$ flux quanta in ${\bf A}_0$. Thus we reobtain the result
already discussed in Sec.~\ref{sec.intro}: Dislocations each
acquire $\pm1/2$ magnetic vortex.

\section{Classical model for\\ the skyrmion lattice}
\label{sec.class}

In the present section we formulate a simple classical model for
the interacting skyrmion system. We use this model to obtain
(a) the classical ground state of the skyrmion lattice for a wide
range of values of the skyrmion density and the
magnetic interaction strength, and (b) the spectrum of low-lying
collective excitations. This model should represent the physics of the
real skyrmion lattice at least qualitatively, and even
quantitatively at low density.
This section is meant to illustrate some of the properties to be
expected for the skyrmion lattice without introducing irrelevant
technical complications. Furthermore, we wish to motivate the
choice of lattice types studied on a more general level in
Sec.~\ref{sec.KT}.

\subsection{Model}
\label{sec.mod}

The main idea is to take the correct limit of the
two-skyrmion interaction at large distances and treat the skyrmion
system as a classical gas of point particles having this
interaction at all separations. We thus keep only the respective
leading-order terms for large separations of both the interaction
contribution independent of the $XY$ degree of freedom and of the
contribution depending on this degree of freedom, and
we neglect \mbox{three-,} \mbox{four-,} etc.\ body interactions.
This model should be valid at low skyrmion densities.

We start from the classical non-linear sigma model for the
magnetization,\cite{Fsigma,Auerbach} which has been successfully
applied to quantum Hall ferromagnets.\cite{Sondhi,Moon}
Abolfath {\it et al}.\cite{Abol} have recently discussed the
applicability of this classical field theory and compared its
predictions with microscopic results.
The magnetization is represented by a normalized three-component
vector field ${\bf m}({\bf r})$. The relevant terms in the
Lagrangian read
\begin{eqnarray}
L & = & \frac{\hbar}{4\pi\ell^2} \int d^2r\, {\bf A}[{\bf m}]
    \cdot\partial_t{\bf m}
- \frac{\rho_s}{2} \int d^2r\, (\partial_i m_\mu)(\partial_i m_\mu)
  \nonumber \\
& & {}+ \frac{\rho g^\ast \mu_B B}{2} \int d^2r\, m_3({\bf r})
  \nonumber \\
& & {}- \frac{e^2}{2\epsilon} \int d^2r\,d^2r'\, \delta\rho({\bf r})
    \frac1{|{\bf r-r'}|} \delta\rho({\bf r'}) ,
\label{1L1}
\end{eqnarray}
where ${\bf A}[{\bf m}]$ is the vector potential of a magnetic
mo\-no\-pole at the origin in spin ({\bf m}) space,\cite{QHferro}
$\partial_t$ is a time derivative, and
\begin{equation}
\delta\rho \equiv -\frac1{8\pi} \epsilon_{ij} {\bf m}\cdot
  (\partial_i{\bf m}\times\partial_j{\bf m})
\end{equation}
is the topological (Pontryagin) density.
Greek indices always run over three values and latin ones over two.
The first term is the usual Berry phase, and the other three
stem from exchange, Zeeman, and Coulomb interaction, respectively.
The Coulomb term reflects the fact that skyrmions carry electrical
charge.

In the absence of Zeeman and Coulomb interactions, the ground-state
solution for a single skyrmion is known analytically.\cite{BP}
It is scale invariant and for
large distances $r=|{\bf r}|$ from the skyrmion center the in-plane
components of ${\bf m}$ fall off as $m_j\propto r_j/r^2$. Rotation
around the $z$ axis gives a {\it different\/} ground-state
solution, reflecting the completely broken SO(3) symmetry.

Switching on the Zeeman interaction, scale invariance is broken since
the Zeeman term prefers a small skyrmion. Far away from the
skyrmion center we can expand the exchange and Zeeman terms in the
Lagrangian, Eq.~(\ref{1L1}), up to second order in the small
in-plane components $m_j$ of the magnetization,\cite{HS}
\begin{equation}
E \cong \int d^2r\, \left[
  \frac{\rho_s}{2} (\partial_i m_j)(\partial_i m_j)
  + \frac{\rho g^\ast \mu_B B}{2}\, m_jm_j \right] .
\end{equation}
The resulting Euler-Lagrange equation
\begin{equation}
-\partial_i\partial_i m_j + \frac{\rho g^\ast \mu_B B}{2}\,
  m_j = 0
\end{equation}
has a vortex solution with\cite{HS}
\begin{equation}
m_j \propto \frac{r_j}{r^{3/2}}\,e^{-\kappa r}
\label{1m1}
\end{equation}
for $j=1,2$. This expression is valid for $r\gg 1/\kappa$,
where $\kappa^2 = \rho g^\ast\mu_B B/2\rho_s$.
Again the in-plane components can be rotated through any angle $\phi$.
By inserting ${\bf m}$ into Eq.~(\ref{1L1})
it is seen that the energy-density contributions of
both the exchange and the Zeeman term fall off as
$r^{-1} e^{-2\kappa r}$. Taking the Coulomb interaction into account,
its leading-order contribution to the energy density also behaves like
$r^{-1} e^{-2\kappa r}$. Thus all
three energy contributions are equally relevant at large $r$
and Coulomb interaction does not destroy the
functional form of Eq.~(\ref{1m1}) but does change the
value of $\kappa$.

In calculating the interaction energy we assume that the
two-skyrmion state with one skyrmion at the origin and the
other at ${\bf s}$ is well described by
\begin{equation}
m_j({\bf r}) = m_j^{\text{single}}({\bf r}) + R_{jk}(\phi)
  m_k^{\text{single}}({\bf r-s})
\end{equation}
for $j=1,2$. The component $m_3$ is determined by
$|{\bf m}|=1$. This ansatz only gives errors of higher order in
$e^{-\kappa s}$ for large separations. Here,
\begin{equation}
R_{jk}(\phi) = \left( \begin{array}{rr}
    \cos\phi & \sin\phi \\
   -\sin\phi & \cos\phi
  \end{array} \right)
\end{equation}
rotates the in-plane ${\bf m}$ components
of one of the skyr\-mions through an angle $\phi$.
We find the interaction potential
by inserting ${\bf m}$ into the potential energy part of
Eq.~(\ref{1L1}) and
subtracting the energies of two isolated skyrmions.
We are interested in the limiting form for
large separations $s$.

The exchange contribution to the interaction is the
only one depending on
the angle $\phi$. Using a multipole expansion and integrating
over ${\bf r}$ we find, to leading order, the exchange
contribution
$E_{\text{exch}}\propto \cos\phi\, e^{-\kappa s}/\sqrt{s}$,
where the coefficient of proportionality is positive.
The contribution from the Zeeman term does not depend on
$\phi$ and is exponentially small for large separations.
We neglect it compared to the Coulomb interaction, below,
since we only keep the leading $\phi$ dependent and the leading
$\phi$ independent term.
%%We do not expect a contribution of the same order from
%%the Zeeman term in the limit of large separations since
%%the total number of flipped spins is quantized. Thus the
%%total spin and consequently the total Zeeman energy of
%%two skyrmions will not change at all for sufficiently
%%large separation. [Note that the skyrmion spin is canonically
%%conjugate to the U(1) angle. Fixing the total spin
%%leads to maximum uncertainty in the {\it global\/}
%%angle but does not affect the {\it relative\/} angle $\phi$,
%%which determines the magnetic interaction.]
For $\nu=1$ the leading contribution from the Coulomb interaction is
$e^2/\epsilon s$, where $\epsilon$ is the dielectric
constant of the material. There are higher multipole terms, which
fall off at least as $1/s^3$ and are neglected compared to the
$1/s$ term.

\subsection{Ground states}
\label{sec.stat}

As long as the skyrmion density is small, we expect the
interaction to be dominated by the two-particle large-separation
terms found above. Our approximation for the energy per skyrmion of
a skyrmion lattice is
\begin{equation}
E = E_C + E_{XY}
\label{1Etot}
\end{equation}
with
\begin{eqnarray}
E_C & = & \frac{e^2}{2\epsilon} \sum_{{\bf R}\neq 0} \frac1{R} - E_0
  , \\
E_{XY} & = & \frac{g_{XY}}{2} \sum_{{\bf R}\neq 0}
  \cos(\phi_{\bf R}-\phi_0)\, \frac{e^{-R/\xi_{XY}}}{\sqrt{R}} ,
\label{EXY}
\end{eqnarray}
where ${\bf R}$ runs over all skyrmion positions in the lattice
except ${\bf R}=0$, $E_0$ is an infinite constant from the Coulomb
interaction with the neutralizing background, $g_{XY}>0$ denotes
the strength of the magnetic interaction, $\phi_{\bf R}$ is
the angle of rotation of the skyrmion at ${\bf R}$, and
$\xi_{XY}\equiv 1/\kappa$ is the range of the magnetic interaction.

The long-range Coulomb interaction is not easy to sum over.
The main idea of how to make this summation well-behaved is
due to Ewald\cite{Ewald} and has been successfully applied
to two-dimensional crystals:\cite{BM,FHM,GP} The lattice sum is split
into a rapidly converging part and a long-range
part, which is mapped onto a rapidly converging sum over
the reciprocal lattice. Here, we quote a more general
result,\cite{BM,GP} which will be useful later: If the ${\bf R}$ are
summed over a two-dimensional Bravais lattice then
\begin{eqnarray}
\lefteqn{
e^{i{\bf k}\cdot{\bf s}} \sum_{\bf R}
  \frac{e^{-i{\bf k}\cdot({\bf R+s})}}
  {|{\bf R+s}|} - \frac1{s}
\!=\! \sqrt{n} \sum_{\bf G} e^{i({\bf G+k})\cdot{\bf s}}\, \Phi\!\left(
  \frac{|{\bf G+k}|^2}{4\pi n} \right) } \nonumber \\
& & \qquad {}+ \sqrt{n} \sum_{{\bf R}\neq 0} e^{-i{\bf k}\cdot{\bf R}}
  \, \Phi\!\left( \pi n |{\bf R+s}|^2 \right)
  \qquad\qquad\qquad\qquad\, \nonumber \\
& & \qquad {}+ \sqrt{n}\: \Phi(\pi n s^2) - \frac1{s} ,
\label{1Ewald1}
\end{eqnarray}
where ${\bf G}$ are the reciprocal lattice vectors, $n$ is
the number density, and
$\Phi(x) \equiv \sqrt{\pi/x}\,\text{erfc}(\sqrt{x})$
with the complementary error function $\text{erfc}$.\cite{GR}
Equation (\ref{1Ewald1}) only works for a
Bravais lattice, lattices with a basis need special
consideration. The magnetic structure is irrelevant here,
since the Coulomb interaction does not depend on $\phi_{\bf R}$.

The simple sum over Coulomb interactions is obtained in the limit
${\bf s}\to0$, ${\bf k}\to0$, where the two sums on the right hand
side can be cast into one,
\begin{equation}
E_C = \frac{e^2}{\epsilon} \sum_{{\bf R}\neq 0}
  \frac{\text{erfc}(\sqrt{\pi n} R)}{R}
  - \frac{2e^2 \sqrt{n}}{\epsilon} .
\label{1EC1}
\end{equation}
Now all lattice sums are rapidly converging and we
can calculate the energy accurately. We write the energy per
skyrmion in a dimensionless form,
\begin{eqnarray}
\tilde{E} \equiv \frac{E}{e^2\epsilon^{-1}\sqrt{n}}
& = & \sum_{{\bf r}\neq 0} \frac{\text{erfc}(\sqrt{\pi} r)}{r} - 2
  \nonumber \\
& & {}+ \frac1{2\alpha} \sum_{{\bf r}\neq 0} \cos(\phi_{\bf r}-\phi_0)\,
    \frac{e^{-r/\beta}}{\sqrt{r}}
\label{1E2}
\end{eqnarray}
with
\begin{eqnarray}
{\bf r} & \equiv & \sqrt{n}\,{\bf R} , \\
\alpha & \equiv & e^2 n^{1/4} / \epsilon g_{XY} , \\
\beta & \equiv & \sqrt{n}\,\xi_{XY} ,
\end{eqnarray}
which are all dimensionless. Note that $\beta^2$ is the density
in units derived from the range of the magnetic interaction and
$\beta/\alpha^2 = \xi_{XY}g_{XY}^2\epsilon^2/e^4$ is a measure
of the relative strength of the magnetic interaction and does not
depend on density.
We expect this model to be quantitatively correct for small
densities, $\beta^2\ll 1$.

The classical ground state for given $\alpha$, $\beta$ is determined
by minimizing the energy (\ref{1E2}). Thus we have to compare
$\tilde{E}$ for all reasonable two-dimensional lattice structures,
taking the magnetic order into account. Besides the
triangular lattice with frustrated antiferromagnetic order
and the square lattice with N\'eel order\cite{Slattice} we
have also obtained ground state energies
for the simple rectangular, the centered rectangular, and the oblique
lattice, thereby covering all two-dimensional Bravais
lattices,\cite{BM} all with N\'eel order, and the honeycomb lattice,
which is also bipartite but is not a Bravais lattice.
We cannot strictly exclude the possibility of more complicated
ground states but have not found any other likely candidate.
In the cases of the rectangular and oblique lattices, the lattice is
characterized by one and two, respectively, continuous parameters
in addition to its space group (Bravais lattice type).
For example, the simple rectangular lattice has the anisotropy $\eta$,
defined as the ratio of the lattice constants in the $(10)$ and $(01)$
directions, as an additional parameter. To find the ground state, these
parameters have to be optimized.

Equation (\ref{1EC1}) is not applicable to the honeycomb lattice since
it is not a Bravais lattice. However, its Coulomb energy
$E_C^H$ can be expressed in terms of the triangular-lattice Coulomb
energy $E_C^T$.\cite{CTnotes} Taking
the different densities into account, we find the dimensionless
energy $\tilde{E}_C^H = (1+\sqrt{3})/(2\sqrt{2})\,\tilde{E}_C^T$.
The Coulomb energies of the parameter-free lattices are
$\tilde{E}_C^S = -1.95013$ for the square lattice,
$\tilde{E}_C^T = -1.96052$ for the triangular lattice, and
$\tilde{E}_C^H = -1.89371$ for the honeycomb lattice.
The first two were also found in Ref.~\onlinecite{BM}.

\begin{figure}
\centerline{\epsfxsize 9cm\epsfbox{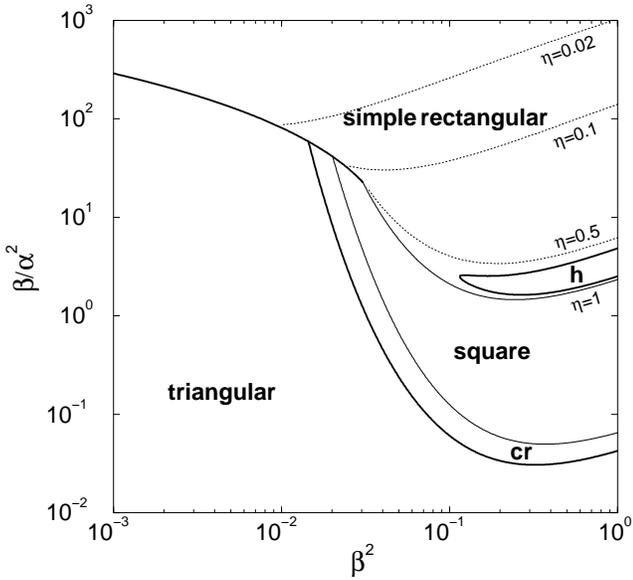}}
\caption{Classical skyrmion lattice phase diagram at $T=0$. Thin
solid lines are continuous phase transitions, whereas heavy lines
are first-order transitions. The honeycomb phase is denoted
by `h' and the centered rectangular phase by `cr.'
The dotted lines in the simple rectangular phase
are lines of constant anisotropy $\eta\le1$.
The employed model is quantitatively correct for
$\beta^2\ll1$. Also, real systems are expected to have
$\beta/\alpha^2\protect\alt 1$.}
\label{fig.PD}
\end{figure}

We map out the ground-state phase diagram in Fig.~\ref{fig.PD}
by following the various transition lines, {\it i.e.}, lines of
equal energy of two lattice types.\cite{CTnotes} We then discard
lines that do not separate two regions with different {\it ground\/}
states. The thin lines denote continuous transitions, whereas
the heavy lines show first-order transitions.\cite{CTnotes}
Recall that our approximation becomes
doubtful for $\beta^2\sim1$, {\it i.e.}, towards the right edge of
the diagram.

The phase diagram, Fig.~\ref{fig.PD}, is quite rich.
For example, there is a region where the ground state is a
honeycomb lattice. In its region of stability, it is even
less frustrated than the square lattice for our model interaction.
In the upper left corner we find a very anisotropic ground state
consisting of widely separated chains of skyrmions.
Another interesting feature is the
critical point on the square--simple rectangular line.
Probably more relevant for real systems is the appearance of a
centered rectangular phase (a square lattice stretched
along the diagonal) everywhere between the triangular and square
lattices. It should be possible to experimentally see this
two-step transition upon varrying $\nu$.
Real skyrmion systems probably live in the lower part of the phase
diagram since the magnetic interaction cannot be made arbitrarily large
in experiment.
The parameter $\beta/\alpha^2$ can be increased by reducing the
Zeeman interaction and thereby increasing the skyrmion size $\xi_{XY}$.
Experimentally, this can be done by applying hydrodynamic pressure.
It is easier to
increase the Zeeman interaction, reducing $\beta/\alpha^2$, by
applying a in-plane magnetic field component. We roughly estimate that
$\beta/\alpha^2$ is smaller than unity in real systems.
%% NEW
The transition lines show an upturn to larger magnetic interactions
at the right edge of the phase diagram, Fig.~\ref{fig.PD}. Although
this may be an artifact of our approximation $\beta^2\ll 1$, it is
interesting to note that a similar reentrance of a triangular phase
is found in Ref.~\onlinecite{Rao}.
%% END NEW

The phase diagram is rather robust against changes
in the exact form of the magnetic interaction. For example, the
phase diagram for a simple exponential magnetic
interaction is qualitatively identical to Fig.~\ref{fig.PD}. This
robustness indicates that the errors made by neglecting higher-order
terms in the magnetic interaction are typically small.

%% NEW
Rao {\it et al}.\cite{Rao} use a variational classical non-linear
sigma model approach to find the classical ground states of the
skyrmion lattice. They only consider the square and triangular
lattices and consequently do not find the other phases, in particular
the centered rectangular lattice. Their method has the advantage
that the skyrmion size is optimized for given density and lattice type.
The single skyrmion magnetization used in Ref.~\onlinecite{Rao} does not
approach the correct limit at large distances but according to the
above argument this should not change the results qualitatively.
However, at low density a {\it ferromagnetically\/} ordered triangular
lattice is found\cite{Rao} which appears to be inconsistent with the
large separation limit of the exchange interaction, Eq.~(\ref{EXY}).
%% END NEW

\subsection{Dynamics}
\label{sec.dyn}

We now briefly turn to the low-energy collective excitations of the
skyrmion lattice. As noted above, the U(1) degree of freedom leads
to the appearance of a gapless $XY$ spin wave mode, whereas
displacements of the skyrmions lead to a magnetophonon mode.
The usual ferromagnetic spin wave mode is gapped at the Zeeman
energy. This mode is expected to mix with the $XY$ mode,
except at ${\bf k}=0$. This effect cannot be reproduced
by the present model where the $S^z$ spin components are completely
integrated out. Thus the magnon dispersion is only reliable for
the long-wavelength acoustical modes.

We denote the displacements of skyrmions from
their ground state positions by ${\bf u}=(u_1,u_2)$ and the
deviation of the angle $\phi$ from its ground state by $u_0$.
Then we expand the potential energy Eq.~(\ref{1Etot}) up to second
order in $u_\mu$.
To describe dynamics we also have to know the leading time-derivative
terms in the Lagrangian. The term for displacements can be
derived from the original Lagrangian, Eq.~(\ref{1L1}).\cite{Stone,HS}
In the limit of vanishing Landau level mixing the skyrmion mass
vanishes so that the Berry phase term is the only relevant one.
One does not normally find a second-order time derivative term in
spin dynamics, but Hartree-Fock calculations\cite{Cote97} clearly show
that $u_0$ obtains a mass, or rather a moment of inertia $I$.
This can be understood as arising from having integrated out all the
short-wavelength spin fluctuations in order to obtain an effective
action for the collective coordinate $u_0$.
There is also a Berry phase term associated with $u_0$
but it is a total time derivative and thus irrelevant at
the classical level. The kinetic terms in the Lagrangian are thus
\begin{equation}
T = \frac{I}{2} \sum_{{\bf R},n} \dot{u}_0^n({\bf R})
    \dot{u}_0^n({\bf R})
  + \frac{\eta}{2} \sum_{{\bf R},n} \epsilon_{ij} u_i^n({\bf R})
    \dot{u}_j^n({\bf R}) .
\label{1Lkin1}
\end{equation}
Here, ${\bf R}$ is a Bravais lattice vector of the {\it magnetic\/}
lattice, the superscript $n=0,1,\ldots$ selects one skyrmion of the
lattice basis, $I$ is the moment of inertia of $u_0^n({\bf R})$,
and the coefficient $\eta$ is\cite{Stone,HS} $\eta = e^\ast B$, where
$e^\ast = \pm e$ is the skyrmion charge. Deriving the Euler-Lagrange
equations and making a plane-wave ansatz,
\begin{equation}
u_\mu^n({\bf R}) = A_\mu^n e^{-i({\bf k}\cdot{\bf R}-\omega t)} ,
\label{1ans1}
\end{equation}
we find the equations of motion\cite{CTnotes}
\begin{eqnarray}
0 & = & \left\{ \begin{array}{ll}
    I \omega^2 A_0^n & \mbox{for $\mu=0$} \\
    -i\eta \omega \epsilon_{\mu j} A_j^n & \mbox{for $\mu\neq 0$}
  \end{array} \right\} \nonumber \\
& & {}- \sum_{n'} \left[ K_{\mu\nu}^{nn'}(0) A_\nu^n
    - K_{\mu\nu}^{nn'}({\bf k}) A_\nu^{n'} \right]
\end{eqnarray}
with the dynamical matrix
\begin{eqnarray}
\lefteqn{
K_{\mu\nu}^{nn'}({\bf k}) \equiv \sum_{\bf R}
  e^{-i{\bf k}\cdot{\bf R}}\:
  \frac{\partial^2}{\partial s_\mu \partial s_\nu}\, }
  \nonumber \\
& & \qquad \times \left.
  E({\bf R}+{\bf c}^n-{\bf c}^{n'}\!+{\bf s},
    \phi^n-\phi^{n'}\!+s_0) \right|_{{\bf s}=0, s_0=0} .
\end{eqnarray}
Here, $E$ is the potential energy per skyrmion, Eq.~(\ref{1Etot}),
${\bf c}^n$ denotes the position of skyrmion $n$ within the
unit cell, and $\phi^n$ is its $XY$ angle.

We can now address the question of mixing of magnetophonon
and $XY$ magnon modes.
Matrix elements mixing displacements and rotations of the $XY$ angle
($\mu=0$ and $\nu\neq0$ or {\it vice versa\/})
stem from the magnetic interaction alone and
contain a first-order derivative of $\cos(\phi^n-\phi^0)$, {\it i.e.},
$\sin(\phi^n-\phi^0)$, as a factor. Thus for any lattice with
$\phi^n\in\{0,\pi\}$ (square, rectangular, oblique, honeycomb)
these matrix elements vanish. Of the
lattices considered above, only the triangular
shows any mixing of $XY$ magnons and magnetophonons in our model.

\begin{figure}
\centerline{\epsfxsize 9cm\epsfbox{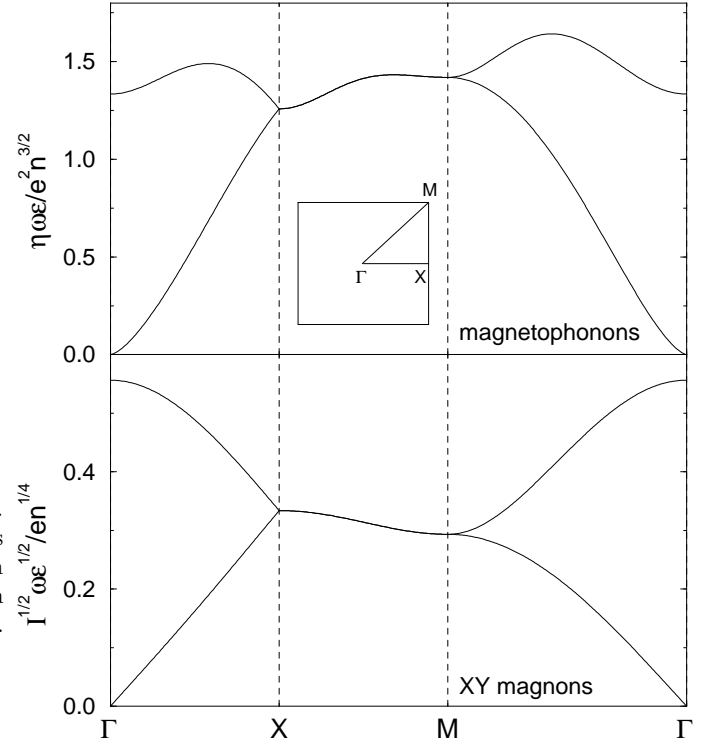}}
\caption{Dispersion of magnetophonons and $XY$ magnons along
high-symmetry directions in the magnetic Brillouin zone of the square
skyrmion lattice. The calculation was done for
$\beta^2=0.1$ and $\beta/\alpha^2=1$.}
\label{fig.dyn}
\end{figure}

Here we only show results for the square lattice. The contribution
from the magnetic interaction can be summed directly, whereas for the
Coulomb interaction we need the second derivatives of
Eq.~(\ref{1Ewald1}). The magnetophonon and
$XY$ magnon dispersions for $\beta^2=0.1$
and $\beta/\alpha^2=1$ are shown along directions of high
symmetry in the magnetic Brillouin zone in Fig.~\ref{fig.dyn}.
There are only two instead of four magnetophonon modes in the
magnetic (folded-back) zone since
$u_1^n$ and $u_2^n$ are canonically conjugate and thus do
not lead to independent modes.
The dispersion of the lower magnetophonon mode for small ${\bf k}$
is indeed of the form $\omega\propto k^{3/2}$, whereas the
magnon mode is linear. Note that the optical magnetophonon branch
shows a minimum at the zone center, a feature previously seen in
Hartree-Fock calculations.\cite{Cote97,remCote97} We see that our
simple model reproduces the main qualitative features of the
collective mode spectrum and conclude that it captures the essential
physics.

The dispersion relations can be used to reintroduce at least
part of the quantum effects into our model
by taking the zero-point energy $\hbar\omega/2$ for all modes into
account.\cite{BM} This additional energy may favor certain lattice
types and thus shift the transition lines. It may also lead to quantum
melting. Furthermore, we could extract low-temperature thermodynamic
properties, such as the free energy, and with its help study
structural phase transitions at finite temperatures.
The harmonic approximation is, in
principle, inconsistent with the description of melting, be it
quantum or thermal, which
is intrinsically non-linear. Nevertheless, estimates
of melting temperatures could be found by calculating the
amplitude of vibrations in this approximation
and using a Lindemann-type criterion.\cite{GP} In particular, we
expect quantum melting due to soft modes in the vicinity
of the continuous transition lines, {\it e.g.}, the one between
square and centered rectangular lattices.

The magnon and magnetophonon dispersions have also been studied by
Green {\it et al}.\cite{Green} This work is not easily comparable
to real systems since the authors assume that the skyrmion shape
is little affected by Zeeman and Coulomb interactions, which
yields an incorrect expression for the interaction at large
distances.
Furthermore, they introduce a mass term into the displacement
equation of motion and their magnetophonon frequencies are
inversely proportional to the mass, which seems doubtful since
the mass vanishes for vanishing Landau level mixing
so that the physics is dominated by the Berry phase term. Two lattice
types are considered in Ref.~\onlinecite{Green}: what we call
the square lattice and the centered rectangular lattice,
the latter with the angle between primitive lattice vectors fixed to
$\pi/3$. The latter is structurally identical to the triangular
lattice but has only two magnetic sublattices. 
The authors dispute the existence of a frustrated triangular
phase\cite{Slattice} but since they
do not consider a triangular lattice with a three-skyrmion basis
it is clear that they cannot find it.

If we go to higher temperatures, the harmonic approximation breaks
down. In particular, topological excitations become important. They
are believed to lead to the ultimate melting of the
skyrmion crystal. In the following section we describe possible
scenarios of melting.

\section{Berezinskii-Kosterlitz-Thouless theory for
skyrmion lattice melting}
\label{sec.KT}

In the present section, the central one of this paper, we discuss
the melting transitions and intermediate phases 
of the lattice types discussed above. We use the framework
of a suitably generalized BKT renormalization group
theory\cite{BKT,Kost74,Halp} since this theory is known to
work well for the simpler problems of (i) an $XY$ model on a
rigid lattice and (ii) a soft lattice without additional
degrees of freedom.\cite{NH,Young} More specifically,
we employ a Coulomb gas language. BKT theory is a {\it static\/}
theory so that the unconventional kinetic terms in the skyrmion
Lagrangian, see, {\it e.g.}, Eq.~(\ref{1Lkin1}), do not affect the
results. We stress that this approach does not make
reference to any special model for the interacting skyrmion system.
There is one important caveat: The theory of Refs.~\onlinecite{NH} and
\onlinecite{Young} describes the system well
{\it if\/} it shows BKT melting but does not say whether it actually
does. The upper or both BKT melting transitions may be preempted by
a first-order transition.

\subsection{Multiple-charge Coulomb gas}

We first introduce a general model, which contains all
relevant skyrmion lattices as special cases. This model
is a two-dimensional continuum Coulomb gas with more than one species
of particles carrying more than one charge.
The dislocations and vortices are treated as classical point
particles with logarithmic interactions. From continuum elasticity
theory on finds that for the triangular lattice
the interaction between two dislocations with Burger's vectors
${\bf b}_1$ and ${\bf b}_2$ and separation vector ${\bf r}$
is proportional to\cite{Nabarro,NH,Young,FHM}
\begin{equation}
-{\bf b}_1\cdot{\bf b}_2\,\ln\frac{|{\bf r}|}{\tau}
  + \frac{({\bf b}_1\cdot{\bf r})({\bf b}_2\cdot{\bf r})}
    {{\bf r}^2} ,
\label{dislint}
\end{equation}
where the length scale $\tau$ is given by the lattice spacing.
For less symmetric lattice types this expression is not strictly
correct but the leading, logarithmic term is always
present.\cite{Young} We here only keep the logarithmic term. The
qualitative behavior and the universal jump in the stiffness
are known to be unaffected by this.\cite{NH,Young} [However, the
temperature dependence of the correlation length above the
dislocation unbinding transition changes for the triangular
lattice as a result of both the sub-leading term in
Eq.~(\ref{dislint}) and the appearance of triplets of dislocations
with vanishing total Burger's vector.\cite{NH,Young}]

As discussed in
the introduction, dislocations attract partial vortices to
minimize the energy resulting from the mismatch in the
antiferromagnetic order. In a Coulomb gas language we have three
charges: the vortex strength and the $x$ and $y$ components of the
dislocation Burger's vector. These three charges correspond to the
three gauge fields discussed in Sec.~\ref{sec.gauge}.
A similar description can be used for the
possible upper melting transition from a liquid crystal to an
isotropic fluid. This transition is thought to be due to
unbinding pairs of disclinations, which may again bind
partial vortices.

The model is defined as follows:
There are $N$ species of particles, counted by $n=1,\ldots,N$,
which carry $M$ charges $q_n^1,\ldots,q_n^M$.
Each particle has an antiparticle with all charges inverted,
$q_{\overline{n}}^m = -q_n^m$,
where we use the notation $\overline{n}$ for the species
of the antiparticle.
The charges interact via the two-dimensional logarithmic Coulomb
potential. The interaction between two particles
of species $n$ and $n'$ at positions ${\bf r}$ and ${\bf r}'$ is then
\begin{equation}
V = -\sum_{m=1}^M q_n^m q_{n'}^m\, \ln\frac{|{\bf r}-{\bf r}'|}{\tau}
  .
\end{equation}
The charges have units of a square root
of energy---the strength of the interaction
is contained in $q_n^m$.
Furthermore, we assume (i) that only a particle and its
antiparticle can form a pair that is neutral with respect
to {\it all\/} charges and (ii) that the whole system
is neutral with respect to all charges. The
number of $n$ particles and $\overline{n}$ antiparticles
is then equal. Restriction (i) is in fact not crucial but
simplifies the argumentation.

The usual Coulomb gas model for vortices in the $XY$ model
or a superfluid film is the case $N=1$, $M=1$. Melting
of a square lattice without additional degrees of freedom
can be described by an $N=2$, $M=2$ Coulomb gas, which reduces to two
independent, identical $N=1$, $M=1$ models since
dislocations with Burger's vectors
along the $x$ and $y$ axes, respectively, do not
interact. Tupitsyn {\it et al}.\cite{Tup} have
considered an $N=2$, $M=2$ model for merons in a double-layer
quantum Hall system, but with a $1/r$ interaction
for one of the charges.

For the square skyrmion lattice we have $M=3$ charges and
$N=5$ particle species. The charges
correspond to the vortex strength ($q_n^1$), the $x$ component of
the Burger's vector ($q_n^2$), and its $y$ component ($q_n^3$).
The charges of the particles are given in table \ref{tabS}.
This table reflects the fact that dislocations bind
$+1/2$ or $-1/2$ vortex.

We apply Kosterlitz' renormalization group
for\-mu\-la\-tion\cite{Kost74} of BKT theory to the general
$N$, $M$ model. This approach has the advantage of being
more rigorous and mathematically more transparent than, {\it e.g.}, the
original self-consistent screening approach.\cite{BKT,Halp}
The renormalization procedure of Ref.~\onlinecite{Kost74} consists
of two steps: First, the smallest neutral pair is integrated
out, and then the length scale $\tau$, the size
of the smallest pairs, is rescaled. The procedure for the
multiple-charge Coulomb gas is similar to the original
case\cite{Kost74} and is not given in detail.

The main approximation of the original theory\cite{Kost74} is
that the two particles with smallest separation are assumed to
{\it always\/} form a neutral pair. This is reasonable
since two particles which do not form a neutral
pair have the same charge and thus repel each other.
In our case this is not generally true: two particles which
are not a neutral pair can even attract each other and
form a non-neutral bound state. However, this arrangement
will attract other particles until it is totally neutral. The
approximation that the smallest pair is neutral is thus equivalent
to neglecting neutral arrangements of more than two particles.
The same problem arises for the triangular lattice without $XY$
degree of freedom, where three elementary dislocations can have
vanishing total Burger's vector.\cite{Young}
Depending on the $q_n^m$ there can be neutral triplets, {\it e.g.},
$(1\overline{2}3)$ in the above example,
whereas in the $N=1$ case the simplest neutral arrangements except
for pairs are quartets. Therefore the first neglected
term in the renormalization group equations is of fourth order
in the particle fugacities for both the
$XY$ model on a rigid lattice\cite{y4term} and the melting square
lattice without $XY$ degree of freedom, whereas it is of
third order for the square skyrmion lattice.

The grand canonical partition function of the multiple-charge Coulomb
gas is
\begin{eqnarray}
Z & = & \sum_{{\cal N}_1,\ldots,{\cal N}_N} \prod_{n=1}^N \left[
  \frac1{({\cal N}_n!)^2} \left(\frac{y_n}{\tau^2}\right)^{2{\cal N}_n}
  \right] \int_{D_1(\tau)} \!\!\! d^2r_1 \ldots  \nonumber \\
& & \int_{D_{2\cal N}(\tau)}
  \!\!\! d^2r_{2\cal N}\,
  \exp\!\left(\! +\frac{\beta}{2} \sum_{i\neq j} \sum_{m=1}^M
    q_{n_i}^m q_{n_j}^m \ln\frac{|{\bf r}_i-{\bf r}_j|}{\tau}
  \right)\! , \nonumber \\
& & {}
\label{Z0}
\end{eqnarray}
where ${\cal N}_n$ is the number of particles of species $n$ (there
is the same number of $\overline{n}$ antiparticles),
${\cal N} = \sum_n {\cal N}_n$, the $y_n$ are fugacities,
the ranges of integration $D_i(\tau)$ comprise the whole plane
but exclude configurations with two particles closer than $\tau$, and
the double sum $\sum_{i\neq j}$ runs over all $2{\cal N}$ particles and
antiparticles. Pairs of size in $[\tau,\tau+d\tau)$ are integrated
out according to
\begin{eqnarray}
\lefteqn{
\prod_{i=1}^{2\cal N} \int_{D_i(\tau)}\!\! d^2r_i \cong
  \prod_{i=1}^{2\cal N} \int_{D_i(\tau+d\tau)}\!\!\!\!\!\! d^2r_i
  } \nonumber \\
& & \qquad {}+ \frac12 \sum_{i\neq j} \prod_{k\neq i,j}
  \int_{D_k(\tau+d\tau)}
    \!\!\!\! d^2r_k \int_{D'} d^2r_j \nonumber \\
& & \qquad \quad\times \int_{\tau\leq |{\bf r}_i-{\bf r}_j|<\tau+d\tau}
    \!\!\!\!\!\! d^2r_i\; \delta_{n_i,\overline{n}_j} .
\end{eqnarray}
Here, $D'$ consists of the whole plane except for disks of radius
$\tau$ centered at all particles $k\neq i,j$.
The only approximation here is contained in the symbol
$\delta_{n_i,\overline{n}_j}$, which states that only neutral pairs
are integrated out. Applying this prescription to Eq.~(\ref{Z0}) and
rescaling $\tau$ we obtain (cf.~Ref.~\onlinecite{Kost74})
\begin{eqnarray}
Z & \cong & Z_0 \sum_{{\cal N}_1,\ldots,{\cal N}_N} \prod_{n=1}^N
  \Bigg[ \frac1{({\cal N}_n!)^2}
    \left(\frac{y_n}{(\tau+d\tau)^2}\right)^{\!2{\cal N}_n}
  \nonumber \\
& & \quad \times \left(1+[2-\beta/2 \sum_{m=1}^M (q_n^m)^2]
      \frac{d\tau}{\tau}\right)^{\!2{\cal N}_n} \Bigg] \nonumber \\
& & \times \prod_{i=1}^{2\cal N}
  \int_{D_i(\tau+d\tau)}\!\!\!\! d^2r_i \,
  \exp\!\Bigg( +\frac{\beta}{2} \sum_{i\neq j} \Bigg[
    \sum_{m=1}^M q_{n_i}^m q_{n_j}^m \nonumber \\
& & \quad {}- 2\pi^2 \sum_{n=1}^N y_n^2 \frac{d\tau}{\tau} \beta
    \left(\sum_{m=1}^M q_{n_i}^m q_n^m\right) \!\!
    \left(\sum_{m=1}^M q_n^m q_{n_j}^m\right)
  \Bigg] \nonumber \\
& & \times \ln\frac{|{\bf r}_i-{\bf r}_j|}{\tau+d\tau} \Bigg) .
\end{eqnarray}
Except for the irrelevant constant $Z_0$, this partition function
is identical to the original one if we replace
\begin{eqnarray}
y_n & \to & \left[1+\left(2-\frac{\beta}{2}
  \sum_{m=1}^M (q_n^m)^2\right)\frac{d\tau}{\tau} \right] y_n , \\
\sum_{m=1}^M q_n^m q_{n'}^m & \to & \sum_{m=1}^M q_n^m q_{n'}^m
  - 2\pi^2 \sum_{n''=1}^N y_{n''}^2 \beta \nonumber \\
& & \times \left(\sum_{m=1}^M q_n^m q_{n''}^m\right) \!\!
    \left(\sum_{m=1}^M q_{n''}^m q_{n'}^m\right) \frac{d\tau}{\tau} .
\end{eqnarray}
If we define a (symmetric) stiffness tensor
\begin{equation}
K_{nn'} \equiv \frac{\beta\sum_m q_n^m q_{n'}^m}{2\pi}
\end{equation}
and express the scaling relations for $y_n$ and $K_{nn'}$ by
differential equations, we obtain the generalized renormalization
group equations
\begin{eqnarray}
\frac{dy_n^2}{d\ell} & = & 2(2-\pi K_{nn})\, y_n^2 ,
\label{Keq1} \\
\frac{dK_{nn'}}{d\ell} & = & -4\pi^3 \sum_{n''=1}^N y_{n''}^2
  K_{nn''} K_{n''n'} ,
\label{Keq2}
\end{eqnarray}
where $\ell=\ln r/\tau$ is the logarithmic length scale.
The initial conditions for these equations are
\begin{eqnarray}
y_n^2(\ell=0) & = & C_n^2 e^{-2\beta E_n^{\text{core}}} , \\
K_{nn'}(\ell=0) & = & \frac{\beta\sum_m q_n^m(0) q_{n'}^m(0)}{2\pi} ,
\end{eqnarray}
where $C_n$ are constants of the order of unity and
$E_n^{\text{core}}$ are the core energies of one $n$ particle.
Note that the generalized Kosterlitz
equations and the initial conditions reduce to the standard
BKT expressions for $N=1$, $M=1$.

\subsection{Skyrmion lattices}

We first consider the square skyrmion lattice ($N=5$, $M=3$). At first
glance this problem looks rather complicated, since it involves coupled
differential equations in $5$ fugacities and $15$ stiffness constants,
taking the symmetry of $K_{nn'}$ into account.
However, we can simplify the problem considerably by looking for
further symmetries. In a first step we consider the symmetries of
$y_n$ and $K_{nn'}$ at the minimum length scale $\ell=0$ and check
which of these survive for the renormalized quantities at $\ell>0$. From
table \ref{tabS} and
the observation that the core energies of all species of dislocations
should be equal, we see that at $\ell=0$
there are at most seven independent quantities,
\begin{eqnarray}
y_v & \equiv & y_1 , \nonumber \\
y_d & \equiv & y_2 = y_3 = y_4 = y_5 , \nonumber \\
K_v & \equiv & K_{11} , \nonumber \\
K_d & \equiv & K_{22} = K_{33} = K_{44} = K_{55} , \\
K_{vd} & \equiv & K_{12} = -K_{13} = K_{14} = -K_{15} , \nonumber \\
K_{dd} & \equiv & K_{23} = K_{45} , \nonumber \\
K_{dd}' & \equiv & K_{24} = -K_{25} = -K_{34} = K_{35} . \nonumber
\end{eqnarray}
The equations for these quantities
can be read off from Eqs.~(\ref{Keq1}) and (\ref{Keq2}).
It is easy to see that two quantities that are equal
by symmetry at $\ell=0$ remain equal to each other
as we integrate away from $\ell=0$. Hence, the
problem can be reduced to seven coupled equations,
\begin{eqnarray}
{dy_v^2}/{d\ell} & = & 2(2-\pi K_v)\, y_v^2 , \nonumber \\
{dy_d^2}/{d\ell} & = & 2(2-\pi K_d)\, y_d^2 , \nonumber \\
{dK_v}/{d\ell} & = & -4\pi^3 y_v^2 K_v^2 - 16\pi^3 y_d^2 K_{vd}^2 ,
  \nonumber \\
{dK_d}/{d\ell} & = & -4\pi^3 y_v^2 K_{vd}^2 - 4\pi^3 y_d^2
  (K_d^2+K_{dd}^2+2K_{dd}^{\prime 2}) , \nonumber \\
{dK_{vd}}/{d\ell} & = & -4\pi^3 y_v^2 K_vK_{vd}
\label{K2.} \\
& & {}- 4\pi^3 y_d^2 (K_dK_{vd}-K_{dd}K_{vd}+2K_{dd}'K_{vd}) ,
  \nonumber \\
{dK_{dd}}/{d\ell} & = & +4\pi^3 y_v^2 K_{vd}^2 - 8\pi^3 y_d^2
  (K_dK_{dd}-K_{dd}^{\prime 2}) , \nonumber \\
{dK_{dd}'}/{d\ell} & = & -4\pi^3 y_v^2 K_{vd}^2 - 8\pi^3 y_d^2
  (K_dK_{dd}'-K_{dd}K_{dd}') . \nonumber
\end{eqnarray}
In a second step we make an ansatz for the remaining $K$ to
reduce the number of independent quantities further.
We {\it guess\/} that the renormalization
of the interactions can be described in terms of independent screening
of the two charges $q_v$ and $q_d$ alone (we will see below that
this assumption is not correct for all lattice types).
If it were true there would be only two independent
stiffness constants $J_v$ and $J_d$, which we choose so that
\begin{equation}
J_v(0) = \frac{\beta q_v^2(0)}{2\pi} , \qquad
J_d(0) = \frac{\beta q_d^2(0)}{2\pi} .
\end{equation}
Since our ansatz has to work at $\ell=0$ we have
\begin{eqnarray}
K_v & = & J_v , \nonumber \\
K_d & = & J_d + J_v/4 , \nonumber \\
K_{vd} & = & J_v/2 , \\
K_{dd} & = & J_d - J_v/4 , \nonumber \\
K_{dd}' & = & J_v/4 . \nonumber
\end{eqnarray}
Inserting this ansatz into Eqs.~(\ref{K2.}) we obtain four
independent equations,
\begin{eqnarray}
{dy_v^2}/{d\ell} & = & 2(2-\pi J_v)\, y_v^2 ,
\label{K3.1} \\
{dy_d^2}/{d\ell} & = & 2(2-\pi J_d - \pi J_v/4)\, y_d^2 ,
\label{K3.2} \\
{dJ_v}/{d\ell} & = & -4\pi^3 y_v^2 J_v^2 - 4\pi^3 y_d^2 J_v^2 ,
\label{K3.3} \\
{dJ_d}/{d\ell} & = & -8\pi^3 y_d^2 J_d^2,
\label{K3.4}
\end{eqnarray}
and three that are linear combinations of these. Thus
our ansatz is indeed correct. Of course, we could have made this
ansatz directly for the $K_{nn'}$ without the step in between.
We repeat that the leading neglected terms in
Eqs.~(\ref{K3.1})--(\ref{K3.4}) are of third order in $y_{v,d}$.

 From Eqs.~(\ref{K3.3}) and (\ref{K3.4}) we see immediately that if the
dislocations proliferate, $\lim_{\ell\to\infty} y_d^2=\infty$,
both stiffness constants $J_v$ and $J_d$ go to zero for
$\ell\to\infty$. This result reflects the fact that free dislocations
do not only screen the dislocation interaction but also the vortex
interaction since they carry vorticity. On the other hand, free
vortices ($\lim_{\ell\to\infty} y_v^2=\infty$) only lead to $J_v\to 0$
since vortices do not have a non-zero Burger's vector, and,
consequently, Eq.~(\ref{K3.4}) does not contain $y_v^2$.
We now discuss the possible scenarios.

{\it Decoupled transitions}, Fig.~\ref{fig.scen}(a): We start from
low temperatures. At some temperature $T_v$, vortices unbind and
$J_v(\ell\to\infty)$ shows a universal jump from $2/\pi$ to zero.
(We now omit the $\ell$ argument when we refer to the limit
$\ell\to\infty$.) At the same temperature
the {\it effective\/} stiffness of the dislocation interaction,
$J_d+J_v/4$, also shows a jump but at the high-temperature side $T_v^+$
of the jump we still have $J_d+J_v/4 = J_d > 2/\pi$. There is no jump
in $J_d$ alone. Above $T_v$ the system has only short-range magnetic
order but still quasi-long-range translational order. At some
higher temperature $T_d$ the dislocations unbind and $J_d$
jumps from $2/\pi$ to zero. Both transitions show BKT finite size
scaling,\cite{Halp} $J_{v,d}(\ell,T_{v,d}) \cong 2/\pi[1+1/(2\ell)]$.

{\it Vortex driven simultaneous transitions}, Fig.~\ref{fig.scen}(b):
At $T_v$ vortices unbind: $J_v$ jumps from $2/\pi$ to zero.
At $T_v^-$, the effective dislocation stiffness is
$J_d+J_v/4>2/\pi$, while $J_d$ alone is smaller than $2/\pi$.
With $J_v$ jumping to zero, we appear to have $J_d+J_v/4=J_d<2/\pi$
at $T_v^+$, see the cross in Fig.~\ref{fig.scen}(b). However,
from Eq.~(\ref{K3.2}) we see that then
$\lim_{\ell\to\infty}y_d = \infty$, dislocations also
proliferate, and $J_d$ jumps to zero. The physical
reason is that with the vortex interaction screened the
remaining dislocation interaction is suddenly too weak
to bind dislocation pairs so that $T_d=T_v$. We
espect BKT scaling only in $J_v$.

{\it Dislocation driven simultaneous transitions},
Fig.~\ref{fig.scen}(c):
At $T_d$ dislocations unbind: the effective stiffness
$J_d+J_v/4$ jumps from $2/\pi$ to zero. The vortex stiffness
is still large at $T_d^-$, $J_v(T_d^-)>2/\pi$ so that
vortices would unbind at a higher temperature. However,
since $J_d+J_v/4$ vanishes at $T_d^+$, so does $J_v$
so that $T_v=T_d$.
Physically, the vortex interaction is suddenly screened
since the proliferating dislocations carry vorticity.
For this reason the magnetic transition can never take
place at a higher temperature than the melting.
There can be lattice order without magnetic order but
not {\it vice versa}.
We expect BKT scaling in $J_d+J_v/4$ and also find
similar scaling in $J_v$ and $J_d$ separately but with a
non-universal value of $J_{v,d}(T_d^-)\neq 2/\pi$.

\begin{figure}
\centerline{\epsfxsize 7cm\epsfbox{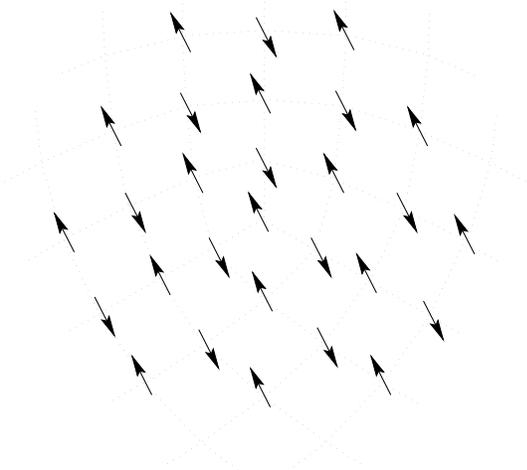}}
\caption{Sketch of a threefold disclination in the square
skyrmion lattice. The disclination does not frustrate the
antiferromagnetic order of the $XY$ spins (arrows).}
\label{fig.disc}
\end{figure}

Above the dislocation unbinding transition at $T_d$ the system
still shows orientational quasi-long-range order, whereas
translational and magnetic order are of short range.
This is the tetratic phase mentioned in the introduction.\cite{NH}
It is characterized by free dislocations, which are in fact
bound pairs of disclinations.\cite{NH} For the square lattice
elementary {\it bare\/} disclinations do not carry vorticity,
as can be seen from Fig.~\ref{fig.disc}.
Disclinations dress with free dislocations, leading to a
logarithmic interaction.\cite{NH} This screening cloud of
dislocations is expected to have vanishing total Burger's vector
and vanishing total vorticity in order to minimize its energy.
Thus dressed disclinations still have zero Burger's vector and
vortex strength and the disclination
unbinding transition corresponds to an $N=1$, $M=1$ model
with the dressed fivefold disclination as the particle and the
dressed threefold disclination as its antiparticle. At
a temperature $T_{\text{disc}}>T_d$ disclination pairs unbind
and the system becomes an isotropic fluid.

We now turn to the other lattice types.
The triangular lattice has dislocations with Burger's vectors
along any of three axes. Dislocations lead to a phase mismatch
of $\pm2\pi/3$ in the frustrated ($120^\circ$) magnetic order
and thus attract $\pm1/3$ vortex.
Consequently, there are six species of dislocations: three
directions of Burger's vectors and two signs of the vorticity.
Dislocations with Burger's vectors in
different directions now interact, the
interaction is proportional to the cosine of the
angle between them,\cite{NH} cf.\ Eq.~(\ref{dislint}).
The charges are given in table \ref{tabT}.
All species of dislocations are equivalent and the final
equations are
\begin{eqnarray}
{dy_v^2}/{d\ell} & = & 2(2-\pi J_v)\, y_v^2 ,
\label{K4.1} \\
{dy_d^2}/{d\ell} & = & 2(2-\pi J_d - \pi J_v/9)\, y_d^2 , \\
{dJ_v}/{d\ell} & = & -4\pi^3 y_v^2 J_v^2 - (8/3)\pi^3 y_d^2 J_v^2 , \\
{dJ_d}/{d\ell} & = & -12\pi^3 y_d^2 J_d^2 .
\label{K4.4}
\end{eqnarray}
Note that there are no neutral triplets in the triangular skyrmion
lattice, as opposed to the usual triangular lattice. Thus the first
omitted terms are of fourth order in the fugacities.\cite{skTxi}
Eqs.~(\ref{K4.1})--(\ref{K4.4}) differ from the square lattice case
only in the
coefficients. The possible melting regimes are thus qualitatively
the same. The liquid crystal phase is hexatic.\cite{NH}
Bare disclinations do carry vorticity ($\pm1/3$) but this fact is
irrelevant for the upper transition since the vorticity part of the
disclination interaction is totally screened by free vortices and
dislocations.

The centered rectangular lattice can be
generated from the square lattice by tilting the
angle $\theta$ between the primitive lattice vectors away from
$\theta=\pi/2$ but keeping their lengths fixed. This tilting
leads to an interaction between dislocations with Burger's vectors
along different primitive vectors. However, all elementary
dislocations still have the same fugacity and effective
interaction with their respective antiparticles since they
are related by reflection symmetry. The charges are
given in table \ref{tabCR}.
For $\theta=\pi/2$ we recover the square lattice.
Our usual ansatz, which assumes that the system can be described
in terms of screening of $q_v$ and $q_d$ alone, fails here. We cannot
reduce the problem to four coupled equations for general
values of $\phi$ but need five:
\begin{eqnarray}
{dy_v^2}/{d\ell} & = & 2(2-\pi J_v)\, y_v^2 , \\
{dy_d^2}/{d\ell} & = & 2(2-\pi J_+/2-\pi J_-/2-\pi J_v/4)\, y_d^2 , \\
{dJ_v}/{d\ell} & = & -4\pi^3 y_v^2 J_v^2 - 4\pi^3 y_d^2 J_v^2 , \\
{dJ_\pm}/{d\ell} & = & -8\pi^3 y_d^2 J_\pm^2
\end{eqnarray}
with the initial conditions for the $J$
\begin{eqnarray}
J_v(0) & = & \beta q_v^2(0) / 2\pi , \\
J_\pm(0) & = & (1\pm\cos\phi)\, \beta q_d^2(0) / 2\pi .
\end{eqnarray}
Nevertheless there can be at most two transitions (apart from
disclination unbinding) since there are only two fugacities. The
possible regimes are the same as for the square lattice if
we replace $J_d$ by $(J_++J_-)/2$. Above the lower melting
transition the system is in a liquid crystal phase with
quasi-long-range orientational order with respect to a
{\it twofold\/} symmetry. This is a two-dimensional {\it nematic\/}
phase. Elementary disclinations do not carry vorticity so that
the upper melting transition is simple.

The simple rectangular lattice is generated from the square lattice
by stretching it in the $(10)$ direction. Dislocations
with Burger's vectors in the $x$ and $y$ direction, respectively,
now have different energies, increasing the number of
independent variables. Dislocations still bind $\pm1/2$ vortex.
There are again $N=5$ particles and $M=3$ charges.
The final renormalization group equations are
\begin{eqnarray}
{dy_v^2}/{d\ell} & = & 2(2-\pi J_v) y_v^2 , \\
{dy_{d1}^2}/{d\ell} & = & 2(2-\pi J_{d1} - \pi J_v/4) y_{d1}^2 , \\
{dy_{d2}^2}/{d\ell} & = & 2(2-\pi J_{d2} - \pi J_v/4) y_{d2}^2 , \\
{dJ_v}/{d\ell} & = & -4\pi^3 y_v^2 J_v^2 - 2\pi^3 y_{d1}^2 J_v^2
  - 2\pi^3 y_{d2}^2 J_v^2 , \\
{dJ_{d1}}/{d\ell} & = & -8\pi^3 y_{d1}^2 J_{d1}^2 , \\
{dJ_{d2}}/{d\ell} & = & -8\pi^3 y_{d2}^2 J_{d2}^2 .
\end{eqnarray}
Since the dislocation energies are different, there are
two distinct dislocation unbinding transitions at $T_{d1}$
and $T_{d2}$ except if they
are driven by vortex unbinding. If there are two structural
transitions, the phase between them has translational quasi-long-range
order in, say, the $x$ direction, but short-range order in
the $y$ direction. It is thus a two-dimensional {\it smectic\/}
phase. There are five possible scenarios, which follow from our
considerations for the square lattice and are not discussed here.
For $T>\max(T_{d1},T_{d2})$ we again have a nematic phase
(twofold rotational symmetry).
Elementary disclinations do not carry vorticity and
the upper melting transition is simple.

The lower melting transition of the honeycomb lattice is trivial in
this context since its dislocations
do not carry vorticity: The structural lattice
has the same basis as the magnetic lattice so that dislocations (with
any Burger's vector) cannot lead to a phase mismatch. Thus
antiferromagnetic quasi-long-range order could persist
in the liquid crystal phase above the dislocation unbinding
transition, which is expected to have unique properties.\cite{CTnotes}
Disclinations carry $\pm1/2$ vortex and any remaining magnetic order
is destroyed at the upper melting transition.

Finally we stress again that these
considerations can only yield the possible sequence of
{\it BKT transitions\/} for any lattice type. This
approach cannot describe other transitions directly,
such as first-order melting transitions\cite{NH} or
first or second-order structural transitions. Structural
transitions in particular could take place
since magnetic disordering leads to a reduced
effective magnetic interaction between skyrmions,
affecting the stability of the various lattice types
in different ways.

In fact, the question arises of whether the square skyrmion lattice
can be stable at all above the magnetic disordering transition.
The answer is affirmative since the magnetic part of the interaction
is of short range: As long as the magnetic BKT correlation length
$\xi_v$, which for $T \agt T_c$ satisfies\cite{BKT,Halp}
\begin{equation}
\frac{\xi_v(T)}{\tau} \cong
  \exp\!\left(\frac{b}{\sqrt{T-T_c}}\right) ,
\label{2xi}
\end{equation}
is much larger than the range of the magnetic interaction, $\xi_{XY}$,
the effect of magnetic disordering on the lattice energies is
negligible. Only when, at a higher temperature, $\xi_v$
becomes comparable to $\xi_{XY}$,
frustration becomes important. In this case we expect the
magnetic interaction to be effectively reduced so that eventually
the triangular lattice becomes favorable.
If at $T=T_d$ still $\xi_v \agt \xi_{XY}$, the square lattice melts
and forms a tetratic phase before a structural transition takes place.
It can even exhibit the upper melting transition without any
structural transition taking place, depending on the non-universal
constant $b$ in Eq.~(\ref{2xi}). The same kind of argument holds for
other lattice and liquid crystal structures.

\section{Conclusions}

\begin{figure}
\centerline{\epsfxsize 9cm\epsfbox{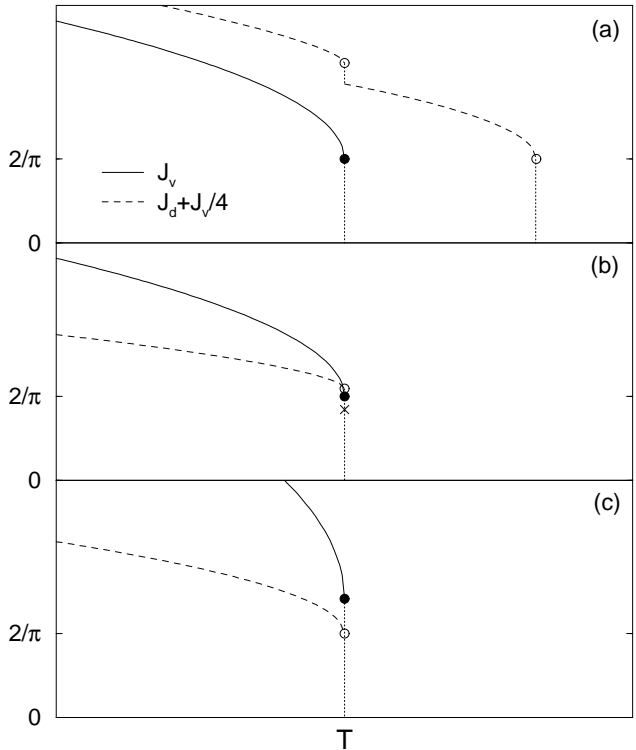}}
\caption{The three possible scenarios close to the lower melting
transition of the square skyrmion lattice: (a) decoupled transitions,
(b) vortex driven simultaneous transitions, and (c) dislocation
driven simultaneous transitions, see text. The vortex stiffness $J_v$
and the effective dislocation stiffness $J_d+J_v/4$ are plotted as
functions of temperature. The graphs have not been obtained
by actual integration but are rather sketches meant to emphasize
the universal features.}
\label{fig.scen}
\end{figure}

As noted above, the lattice types relevant for present
quantum Hall systems are the triangular, centered rectangular,
and square lattices.
A rough estimate of the actual core energies and interactions in the
real quantum Hall system using the model of Sec.~\ref{sec.mod} and
results of Ref.~\onlinecite{FHM}
indicates that the low-density, triangular
lattice typically shows decoupled magnetic and melting transitions,
cf.\ Fig.~\ref{fig.scen}(a). The magnetic skyrmion interaction,
and thus the magnetic stiffness and the vortex energies, are small
since a strong magnetic interaction would make the triangular lattice
unstable. The effective magnetic stiffness is further reduced
by frustration.

On the other hand, the centered rectangular and square lattices
usually show simultaneous transitions, {\it i.e.}, magnetic order
persists up to the lower melting temperature. Experimentally, the
two possible scenarios of Figs.~\ref{fig.scen}(b) and (c) are probably
not easy to distinguish. Note also that vortex driven simultaneous
transitions, Fig.~\ref{fig.scen}(b), require fine tuning of
magnetic and structural stiffnesses so that this scenario is probably
rare.

Quantum fluctuations can affect these results. As noted above they
are expected to lead to quantum melting in the vicinity of the
classical square--centered rectangular line. They should also destroy
magnetic order in the triangular lattice at sufficiently low
densities, where the magnetic interaction becomes exponentially small.
More experiments are needed to test the predictions of this paper.
Sharp structures in the Knight shift, nuclear relaxation
rates, or specific heat as functions of temperature
would be indications for phase transitions of the skyrmion system.
Particularly valuable would be experiments on the electronic
susceptibility $\chi({\bf q},\omega)$
as a function of temperature and filling factor.

In conclusion, we have performed a
Berezinskii-Koster\-litz-Thou\-less renormalization
group study of melting and magnetic disordering in various lattice
geometries in order to understand the behavior of the skyrmion
lattice in quantum Hall ferromagnets at finite temperatures.
The behavior of the skyrmion system is determined by the two
facts that skyrmions (i) are non-collinear magnetic defects
and (ii) carry electrical charge. In the long-wavelength limit
the in-plane magnetization components
can be described by a U(1) ($XY$) degree of freedom
associated with each skyrmion. The XY ``spins''
couple antiferromagnetically and can lead to
antiferromagnetic quasi-long-range order. Dislocations in most
skyrmion lattice types lead to a mismatch in the $XY$ degree of
freedom, which makes the dislocations bind fractional vortices
and leads to coupling of translational
and magnetic excitations. For most
lattice types there are three distinct scenarios for the
lower melting transition: (i) a BKT magnetic disordering
transition
at a lower temperature than BKT melting, (ii) simultaneous
transitions where the magnetic stiffness shows a universal BKT
jump, and (iii) simultaneous transitions where the effective
dislocation stiffness shows a universal jump.

The lattice types we have studied are motivated by the
possible ground states of a simple classical model of the
skyrmion system, which uses the large-separation
limit of their interaction. It shows a surprisingly rich $T=0$ phase
diagram, which suggests that upon increasing the skyrmion density
a frustrated triangular ground state first gives way to a
centered rectangular lattice with N\'eel order and only at higher
density to a square lattice. Quantum melting is expected to take place
in the vicinity of the latter transition.

\acknowledgments

We wish to thank A.H. MacDonald, S.E. Barrett, and S. Sachdev
for valuable discussions.
This work has been supported by NSF DMR-9714055, NSF CDA-9601632,
and NSF DMR-9503814.
C.T. acknowledges support by the Deutsche Forschungsgemeinschaft.
H.A.F. acknowledges a Cottrell Scholar Award of Research Corporation.

\clearpage
%%% FIGURE CAPTIONS

%%% TABLES

\begin{table}
\begin{tabular}{cccc}
  $n$ & $q_n^1$  & $q_n^2$ & $q_n^3$ \\ \hline
  1   & $q_v$    & 0       & 0 \\
  2   & $q_v/2$  & $q_d$   & 0 \\
  3   & $-q_v/2$ & $q_d$   & 0 \\
  4   & $q_v/2$  & 0       & $q_d$ \\
  5   & $-q_v/2$ & 0       & $q_d$
\end{tabular}
\caption{Coulomb gas charges of the square skyrmion lattice.
$q_v$ is the vortex charge and $q_d$ is the dislocation charge.}
\label{tabS}
\end{table}

\begin{table}
\begin{tabular}{cccc}
  $n$ & $q_n^1$  & $q_n^2$           & $q_n^3$ \\ \hline
  1   & $q_v$    & 0                 & 0 \\
  2   & $q_v/3$  & $q_d$             & 0 \\
  3   & $-q_v/3$ & $q_d$             & 0 \\
  4   & $q_v/3$  & $\cos(2\pi/3)q_d$ & $\sin(2\pi/3)q_d$ \\
  5   & $-q_v/3$ & $\cos(2\pi/3)q_d$ & $\sin(2\pi/3)q_d$ \\
  6   & $q_v/3$  & $\cos(2\pi/3)q_d$ & $-\sin(2\pi/3)q_d$ \\
  7   & $-q_v/3$ & $\cos(2\pi/3)q_d$ & $-\sin(2\pi/3)q_d$
\end{tabular}
\caption{Charges for the triangular lattice.}
\label{tabT}
\end{table}

\begin{table}
\begin{tabular}{cccc}
  $n$ & $q_n^1$  & $q_n^2$           & $q_n^3$ \\ \hline
  1   & $q_v$    & 0                 & 0 \\
  2   & $q_v/2$  & $q_d$             & 0 \\
  3   & $-q_v/2$ & $q_d$             & 0 \\
  4   & $q_v/2$  & $\cos(\theta)q_d$ & $\sin(\theta)q_d$ \\
  5   & $-q_v/2$ & $\cos(\theta)q_d$ & $\sin(\theta)q_d$
\end{tabular}
\caption{Charges for the centered rectangular lattice.}
\label{tabCR}
\end{table}


\begin{thebibliography}{99}
\bibitem{magphon}The Lorentz force mixes the transverse acoustical
phonon mode, which has $\omega_T\propto k$, with the
longitudinal plasmon mode with $\omega_L\propto k^{1/2}$.
Hence the frequency of the resulting magnetophonon mode is
$\omega=\omega_T\omega_L/\omega_c\propto k^{3/2}$,
where $\omega_c$ is the cyclotron frequency.
\bibitem{remCote97}The classical dynamics calculations are in a
sense dual to Ref.~\protect\onlinecite{Cote97}: Our model is valid in
the limit of small skyrmion density, whereas the time-dependent
Hartree-Fock calculations work best at high density.
\bibitem{skTxi}Thus the temperature dependence of the correlation
length should not be changed by the mechanism discussed by
Young, Ref.~\protect\onlinecite{Young}.
\end{thebibliography}
\end{document}